\documentclass[twocolumn,twoside,slac_two]{revtex4}
\usepackage{graphicx}
\usepackage{fancyhdr}
\def\fdg{\hbox{$.\!\!^\circ$}}

\begin{document}

\title{Apparent parsec-scale jet opening angles and $\gamma$-ray brightness of active galactic nuclei}

%

\author{A.B. Pushkarev}
\affiliation{Radio Astronomy Laboratory, Crimean Astrophysical Observatory, 98688 Nauchny, Ukraine}
\affiliation{Pulkovo Astronomical Observatory, Pulkovskoe Chaussee 65/1, 196140 St. Petersburg, Russia}
\author{M.L. Lister}
\affiliation{Department of Physics, Purdue University, 525 Northwestern Avenue, West Lafayette, IN 47907, USA}
\author{Y.Y. Kovalev}
\affiliation{Astro Space Center of Lebedev Physical Institute, Profsoyuznaya 86/32, 117997 Moscow, Russia}
\affiliation{Max-Planck-Institut f\"ur Radioastronomie, Auf dem H\"ugel 69, 53121 Bonn, Germany}
\author{T. Savolainen}
\affiliation{Max-Planck-Institut f\"ur Radioastronomie, Auf dem H\"ugel 69, 53121 Bonn, Germany}

\begin{abstract}
We have investigated the differences in apparent opening angles between the parsec-scale jets of
the active galactic nuclei (AGN) detected by the {\it Fermi}\, Large Area Telescope (LAT) during its 
first 24 months of 
operations and those of non-LAT-detected AGN.
We used 15.4 GHz VLBA observations of 215 sources from the 2 cm VLBA MOJAVE program. The apparent
opening angles were determined by analyzing transverse jet profiles from the data in the image 
plane by using stacked images constructed from all available MOJAVE epochs for a given source.
We confirm our earlier result~\cite{MF3} based on the first three months of scientific 
operations of the LAT. The apparent opening angles of $\gamma$-ray bright AGN are preferentially 
larger than those of $\gamma$-ray weak sources, suggesting smaller viewing angles to the $\gamma$-ray
bright AGN. Intrinsic opening angles for BL~Lacs are wider than those in quasars.
\end{abstract}

\maketitle

\thispagestyle{fancy}


\section{INTRODUCTION}
The results from the {\it Fermi}\, Large Area Telescope (LAT) based on the first three months of scientific 
operations have confirmed the findings of EGRET, showing that the extragalactic $\gamma$-ray sky is 
dominated by active galactic nuclei (AGN). Combining the {\it Fermi} and very long baseline interferometry 
(VLBI) observations, a number of AGN radio/$\gamma$-ray connections have been established, namely, that 
the $\gamma$-ray photon flux correlates with the parsec-scale radio flux density \cite{MF1}, the jet of 
the LAT-detected blazars have higher-than-average apparent speeds \cite{MF2}, larger apparent opening 
angles \cite{MF3}, higher Doppler factors \cite{MF4}, and higher median fractional polarization
in their VLBI cores \cite{MF5}. A non-zero time delay 
between 15~GHz radio emission and $\gamma$-ray radiation has been detected \cite{Pushkarev10}.

In this work, we revisit the issue of comparing apparent and intrinsic jet opening angles as well as viewing 
angles for 162 $\gamma$-ray bright (LAT-detected) \cite{2LAC} and 53 weak (non-LAT-detected) AGN using a 
larger sample of 215 sources from the MOJAVE program.

\section{APPARENT OPENING ANGLES}

The apparent opening angle of a jet can be derived either from structure modelfitting in the $uv$ plane, using 
component sizes and their separations from the core, or in the image plane by making transverse jet profiles. 
Both approaches yielded comparable results \cite{MF3}. In this work, we use the second method.
To derive the apparent jet opening angles, we used the 15~GHz naturally weighted MOJAVE VLBA stacked-epoch 
images that were constructed from all available MOJAVE epochs prior to 2011.0. Each stacked-epoch image was obtained 
using the following procedure: (i) all the available images of a given source were restored with the median 
beam, (ii) the images were shifted so that the core component was placed at the phase center, (iii) averaging 
over the images was applied. 

The opening angle of the jet was calculated as the median value of 
\begin{equation}\label{eq:aoa}
\alpha=2\arctan\left(\frac{\sqrt{d^2-b_\varphi^2}}{2r}\right)\,,
\end{equation}
where $d$ is the full width at half maximum (FWHM) of a Gaussian fitted to the transverse jet brightness profile,
$r$ is the distance to the core along the jet axis, $b_\varphi$ is the beam size along the position angle $\varphi$
of the jet-cut, and the quantity $(d^2-b_\varphi^2)^{1/2}$ is the deconvolved FWHM transverse size of the jet.
The direction of the jet axis was determined using the median of the position angles of all jet components over all the 
epochs from model fitting. The slices were taken at 0.1~mas intervals starting from the position of the VLBI core 
and continuing up to the region where the jet either substantially curves or becomes undetectable. The ridge lines 
for 47 sources with notably bending jets were approximated by two straight lines. The opening angle values were 
calculated using only those slices that had a peak of the fitted Gaussian larger than four times the rms noise level 
of the image.

\begin{figure*}[t]
\centering
\includegraphics[angle=-90,width=65mm]{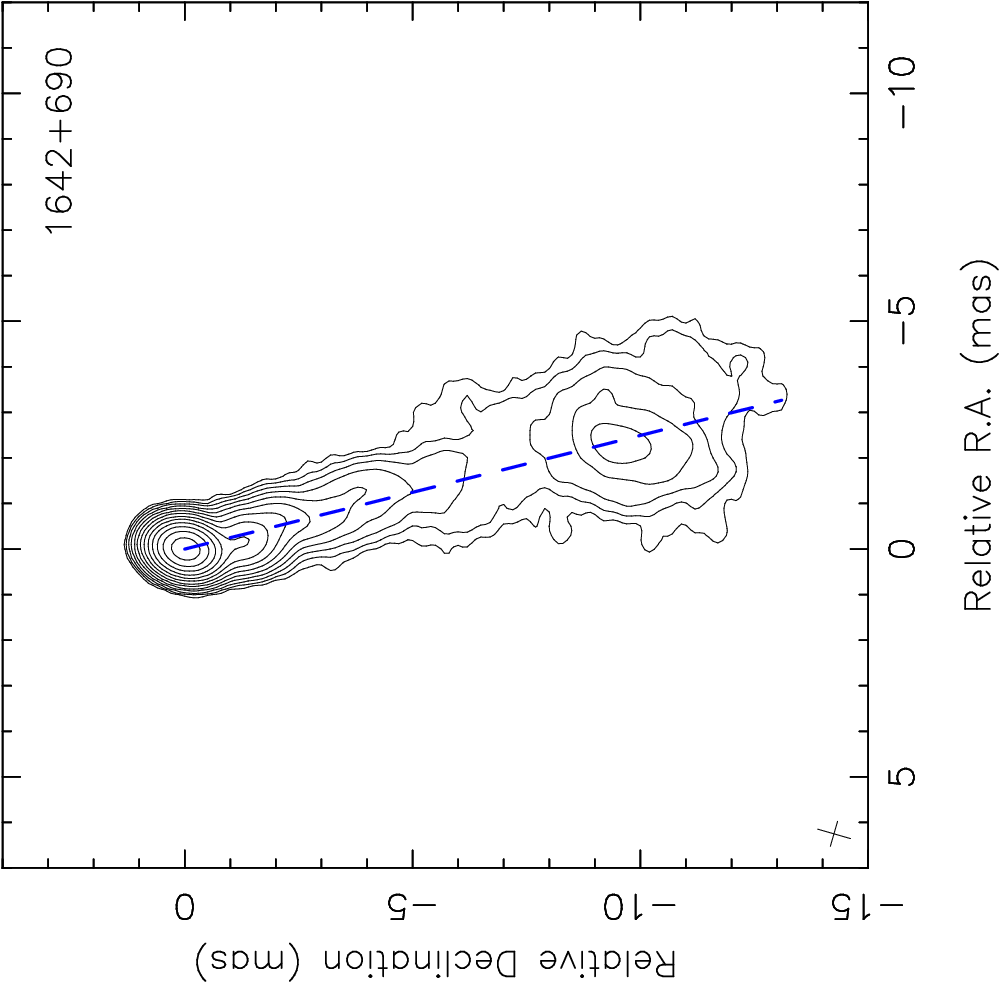}
\includegraphics[angle=-90,width=67mm]{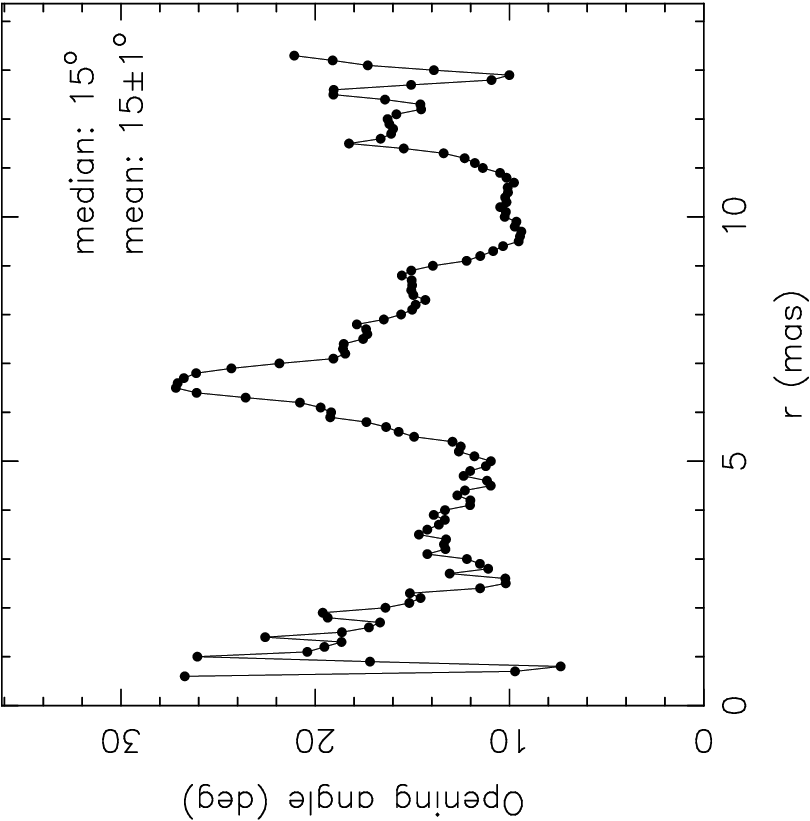}
\caption{{\it Left:} MOJAVE naturally weighted total intensity CLEAN stacked-epoch image of 1642+690 at 15~GHz. 
  The blue dashed line is the jet axis. The contours are plotted at increasing powers of 2, 
  starting from 0.2~mJy~beam$^{-1}$. The peak flux density reaches 483~mJy~beam$^{-1}$.
  The FWHM of the restoring beam is shown as a cross in the lower left corner.  
  {\it Right:} Apparent opening angle of the jet along its axis.} 
\label{f:example}
\end{figure*}

\begin{figure}
\centering
\includegraphics[angle=-90,width=66mm]{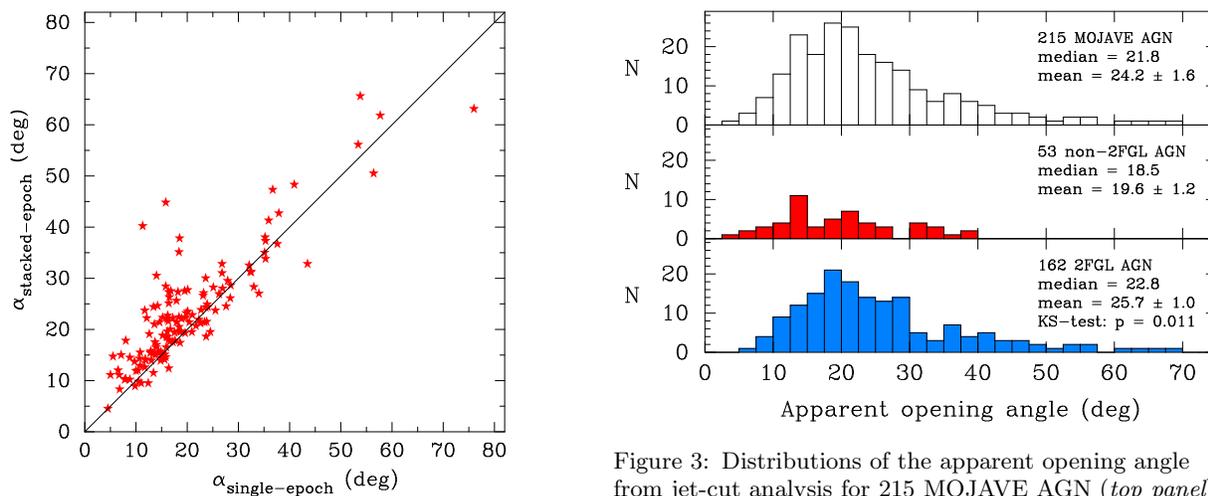}
\caption{Apparent opening angles from single-epoch
         and stacked-epoch images. Solid line is the equality line.}
\label{f:single_vs_stacked}
\end{figure}

\begin{figure}
\centering
\includegraphics[angle=-90,width=79mm]{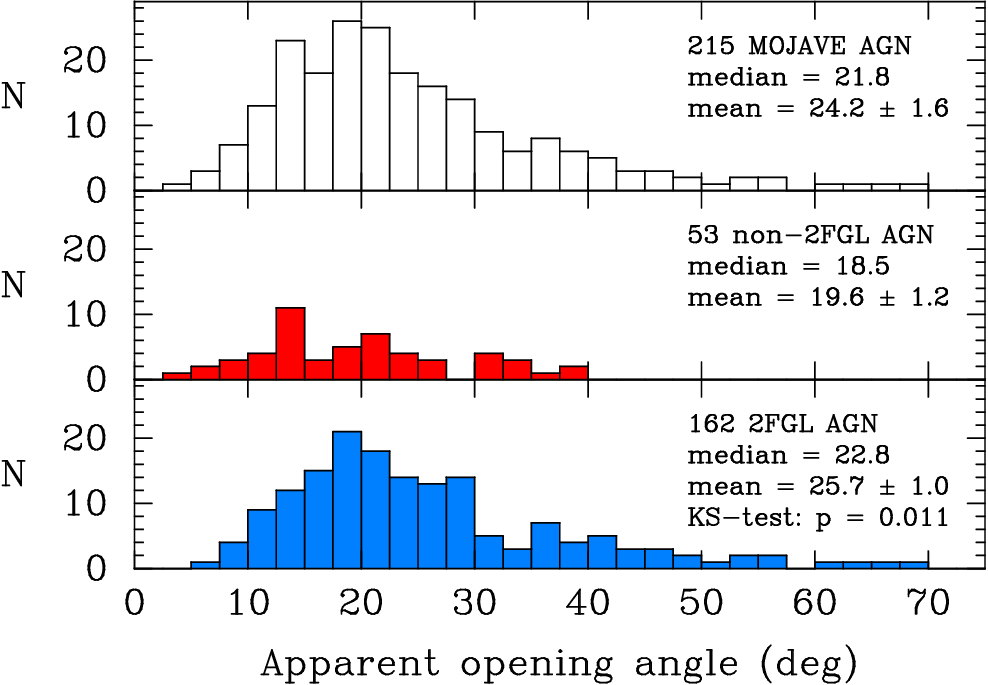}
\caption{Distributions of the apparent opening angle from jet-cut
         analysis for 215 MOJAVE AGN ({\it top panel}), comprising 53 
         non-LAT-detected ({\it middle panel}) and 162 LAT-detected 
         ({\it bottom panel}) sources.}
\label{f:hist}
\end{figure}

\begin{figure*}
\centering
\includegraphics[angle=-90,width=76mm]{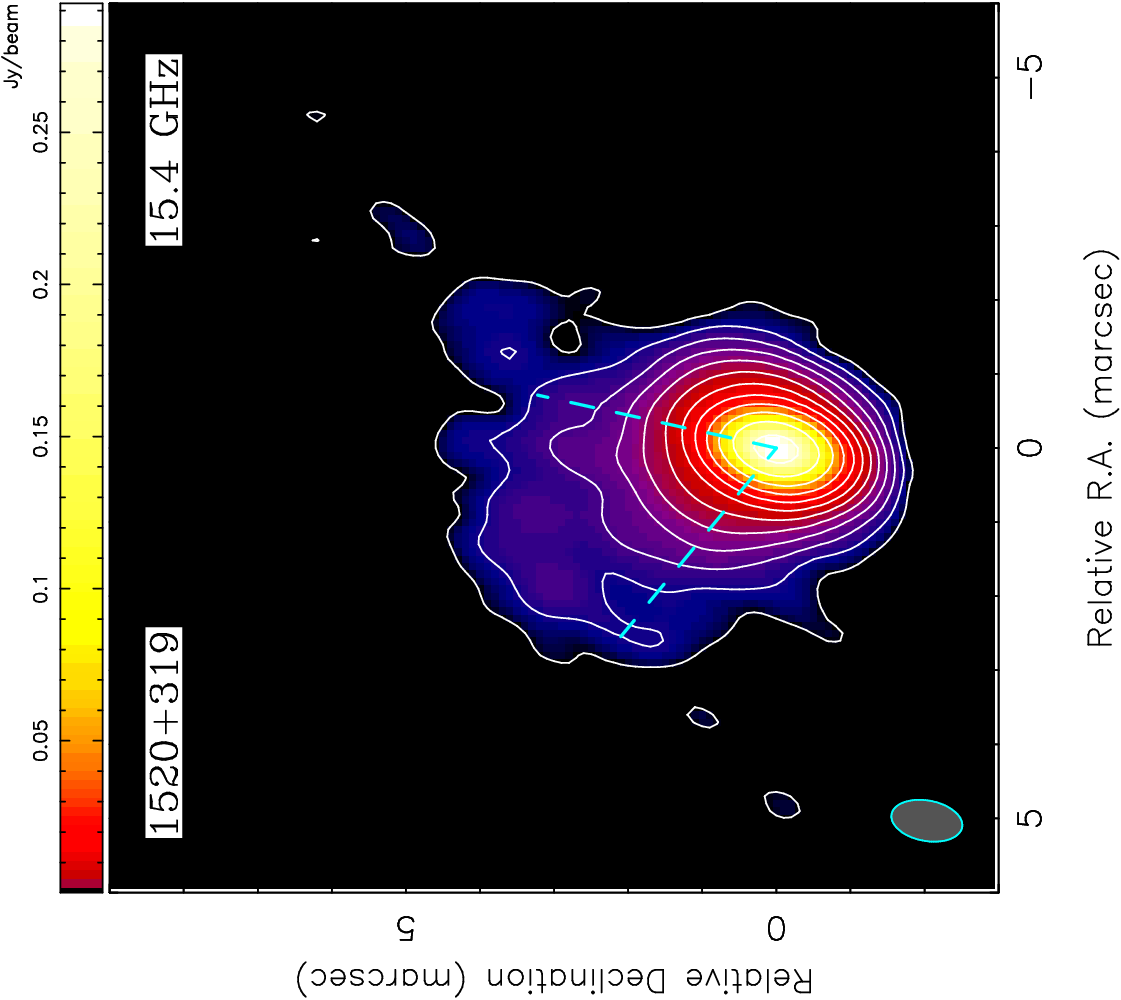}
\includegraphics[angle=-90,width=76mm]{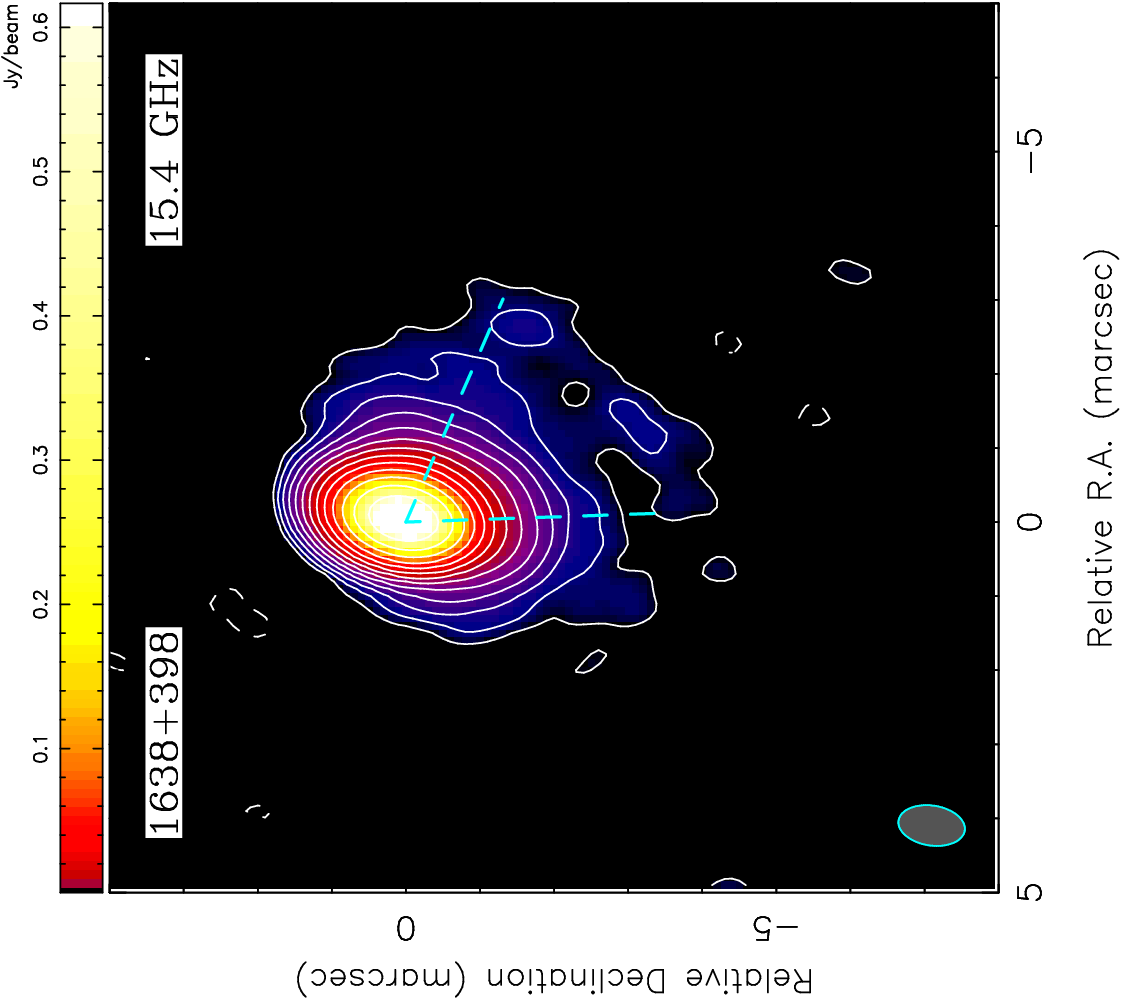}
\caption{MOJAVE images of LAT-detected quasars 1520+319 and 1638+398
         that show the largest apparent opening angles.}
\label{f:maps}
\end{figure*}

In Fig.~\ref{f:example}, the 15~GHz total intensity map of 1642+690 is shown as an example together with opening 
angle of the jet as a function of angular distance to the core. The procedure of stacking all available epochs for 
a given source makes an assessment of the apparent opening angle more accurate as being sensitive to possible changes 
of directions along which different jet features might move over time resulting in a wider apparent opening angle 
than derived from a single-epoch image. In general, the single-epoch opening angles are in good agreement with those 
from the stacked-epoch images Fig.~\ref{f:single_vs_stacked}. In some sources, our measured opening angle was much 
wider due to the presence of low-brightness jet emission that was below the noise level in the single-epoch image.

The distributions of the measured opening angle for 162 LAT-detected and 53 non-LAT-detected sources are shown in 
Fig.~\ref{f:hist}. A Kolmogorov-Smirnov (K-S) test indicates a probability of only $p=0.011$ for these two samples 
being drawn from the same parent population.
In Fig.~\ref{f:maps}, we show MOJAVE VLBI images of two LAT-detected quasars, 1520+319 and 1638+398, with the largest apparent opening angles.

\section{INTRINSIC OPENING ANGLES}

We have derived the values of the viewing angle $\theta$ and the bulk Lorentz factor 
$\Gamma$ for 57 AGN using jet speeds $\beta_\mathrm{app}$ from the MOJAVE kinematic 
analysis~\cite{Lister09} and variability Doppler factors $\delta$ derived from the Mets\"ahovi 
AGN monitoring program~\cite{Hovatta09}. 


The intrinsic opening angles $\alpha_\mathrm{int} = \alpha_\mathrm{app}\,\sin\theta$ were
calculated for the 57 sources (Fig.~\ref{f:lorentz_factor}). A K-S test indicates no significant 
difference ($p=0.42$) between the samples of LAT-detected and non-LAT-detected sources, suggesting 
that the established systematic difference in apparent opening angles is most probably the result
of projection effects, i.e., the $\gamma$-ray bright jets are aligned closer to our line of sight \cite{MF4}.

We also found that BL~Lacs have on-average wider intrinsic opening angles
($2\fdg7\pm0\fdg2$) than those of quasars ($1\fdg4\pm0\fdg1$). The corresponding
distributions are different at confidence level of $>$98.9\% according to the K-S test.
This result is consistent with the higher speeds observed in more powerful outflows of
quasars suggesting a better jet collimation.

\section{VIEWING ANGLES}

\begin{figure}
\centering
\includegraphics[angle=-90,width=75mm]{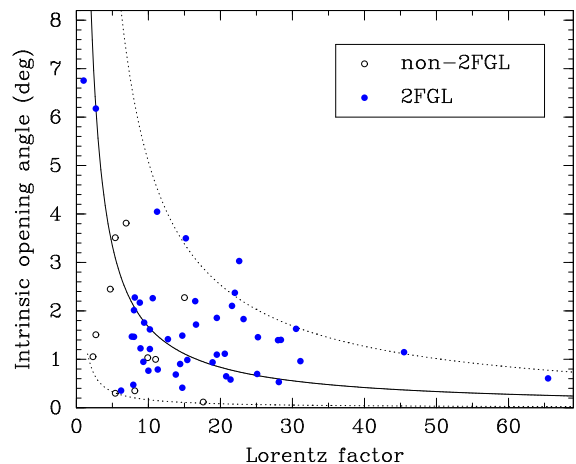}
\caption{Intrinsic opening angle vs. Lorentz factor for 56 jets. 
         The solid line shows the median curve fit with the assumed 
         relation $\alpha_\mathrm{int}=\rho/\Gamma$.}
\label{f:lorentz_factor}
\end{figure}

The viewing angles $\theta$ can be derived from the relation 
$\theta=\mathrm{asin}\,(\alpha_\mathrm{int}/\alpha_\mathrm{app})$. Intrinsic opening angle is inversely 
proportional to Lorentz factor, $\alpha_\mathrm{int}=\rho/\Gamma$ (where $\rho$ is a constant; here $\rho=0.29$~rad), 
as predicted by hydrodynamical \cite{BK79,DM88} and magnetic acceleration models \cite{Komissarov07} of 
relativistic jets, and confirmed by observations (Fig.~\ref{f:lorentz_factor}; see also \cite{Jorstad05,MF3}).

\begin{figure}
\centering
\includegraphics[angle=-90,width=82mm]{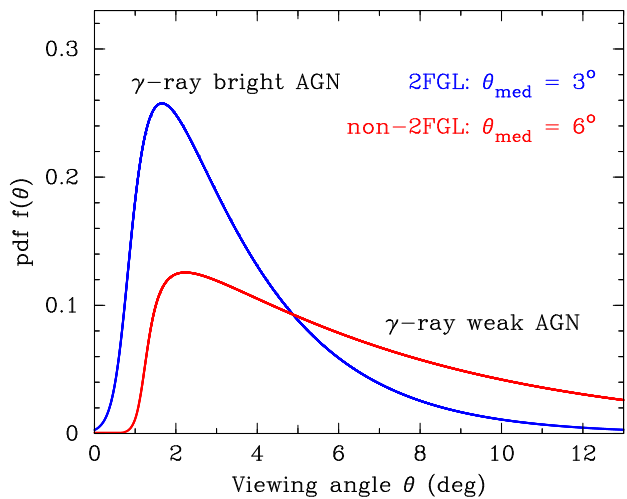}
\caption{Probability density function of viewing angle as derived from the 
         apparent angle and Lorentz factor distributions of $\gamma$-ray 
         bright (blue curve) and $\gamma$-ray weak (red curve) AGN in a flux limited sample.}
\label{f:viewing_angle}
\end{figure}

We simulated the viewing angle distributions
for $\gamma$-ray bright and $\gamma$-ray weak AGN (Fig.~\ref{f:viewing_angle}) by fitting the corresponding
apparent opening angle distributions and assuming flat distributions of $\rho$ within a range of
$[0.1,0.5]$ and Lorentz factor $\Gamma_\mathrm{LAT\_Y}=[7,25]$ and $\Gamma_\mathrm{LAT\_N}=[3,15]$.
Probability density functions as well as empirical density functions for apparent opening angle
distributions and final viewing angle distributions were fitted by the Generalized Lambda Distribution \cite{GLD}.
As seen from Fig.~\ref{f:viewing_angle}, $\gamma$-ray bright AGN tend to have smaller angles to the line of sight
comparing to those of $\gamma$-ray weak AGN, with median values $3^\circ$ and $6^\circ$, respectively.
The probability of observing a jet at $\theta<\alpha_\mathrm{int}/2\simeq0.5^\circ$
is 0.71\%, implying that statistically we expect to have 1 such object in a sample of 162 sources.

\section{Summary}

Using the stacked-epoch images, we have measured the parsec-scale apparent opening angles for the 215 MOJAVE 
sources, 162 of which were detected by the {\it Fermi}\, LAT after the first 24 months of scientific operation. 
The LAT-detected AGN are found to have wider apparent opening angles, with a mean of $25\fdg7\pm1\fdg0$ versus 
$19\fdg6\pm1\fdg2$ 
for the non-LAT-detected sources. All the sources with $\alpha_\mathrm{app}>40^\circ$ are $\gamma$-ray bright.

For 56 sources with known Doppler factors and jet speeds, we derived the intrinsic opening angles. BL~Lacertae
objects have on-average wider opening angles, with a mean of $2\fdg7\pm0\fdg2$ against $1\fdg4\pm0\fdg1$ for quasars.
No significant difference in $\alpha_\mathrm{int}$ is found for $\gamma$-ray bright and weak sources. 
We have also simulated distributions of the viewing angle for the LAT-detected and non-LAT-detected sources and 
confirmed our earlier evidence for $\gamma$-ray bright AGN to have statistically smaller angles to the line of sight.

By analyzing MOJAVE data, we plan to derive Doppler factors for more sources. This will enable us to constrain both
Lorentz factor and viewing angle for these objects, and perform the analysis on a larger sample of sources.

\bigskip 
\begin{acknowledgments}
This research has made use of data from the MOJAVE database that is maintained by the MOJAVE team.
The MOJAVE project is supported under National Science Foundation grant AST-0807860 and NASA 
{\it Fermi} grant NNX08AV67G. YYK was supported in part by the Russian Foundation for Basic Research 
(grant 11-02-00368) and Dynasty Foundation. The VLBA is a facility of the National Science Foundation operated 
by the National Radio Astronomy Observatory under cooperative agreement with Associated Universities, 
Inc.
\end{acknowledgments}

\bigskip 

\end{document}